\documentstyle[epsfig,12pt]{article} 

\newcommand{\be}{\begin{equation}}
\newcommand{\ee}{\end{equation}}
\newcommand{\G}{\Gamma}
\newcommand{\lam}{\lambda}
\newcommand{\g}{\delta g}

\begin{document}
\begin{titlepage}

\begin{flushright}
HET-BROWN-1149
\end{flushright}

\vspace{.4cm}

\begin{center}
\textbf{Back Reaction of Gravitational Radiation on the
Schwarzschild Black Hole}
\end{center}
\begin{center}
L. Raul Abramo$^{1, \dagger}$ and Fabio Finelli$^{2, \ddagger}$ 
\end{center}

\begin{center}
\textit{(1) Physics Department, University of Florida, 
	Gainesville, FL 32611} and \\
\textit{Theoretische Physik, Ludwig-Maximilians-Universit\"{a}t, \\
	Theresienstr. 37, D-80333 M\"{u}nchen, Germany} 

\vspace*{.3cm}

\textit{(2) INFN and Physics Department, University of Bologna, Italy,\\ 
Physics Dept, Brown University, Providence, RI 02912 USA} and \\ 
\textit{Istituto Te.S.R.E./CNR, Via Gobetti 101, 40129 Bologna, Italy}
\end{center}

\vspace*{1cm}

\begin{center}
ABSTRACT
\end{center}

\noindent \hspace*{.2cm} We address some of the issues that appear in
the study of back reaction in Schwarzschild backgrounds. Our main object
is the effective energy-momentum tensor (EEMT) of gravitational
perturbations. It is commonly held that only asymptotically flat or
radiation gauges can be employed for these purposes. We show that the
traditional Regge-Wheeler gauge for the perturbations of the Schwarszchild
metric can also be used for computing physical quantities both at the
horizon and at infinity. In particular, we find that the physically
relevant components of the EEMT of gravitational perturbations have the
same asymptotic behaviour as the stress-energy tensor of a scalar field in
the Schwarzschild background, even though some of the metric components
themselves diverge.

\begin{flushleft}

\vspace*{2cm}

$^\dagger$ e-mail: abramo@theorie.physik.uni-muenchen.de \\
$^\ddagger$ e-mail: finelli@tesre.bo.cnr.it
\end{flushleft}

\end{titlepage}

\section{Introduction}

The study of back reaction of gravitational waves on black hole
spacetimes has been hampered by technical and conceptual difficulties,
from the lack of closed-form expressions for the perturbations to subtle
issues of gauge. The classical
problem of the back reaction of a metric perturbation propagating out of
(or into) a black hole is very interesting, but already complicated
enough, and any progress
in the classic realm would facilitate tremendously the analysis of
quantum effects.

One of the main difficulties in considering the back-reaction of
linearized perturbations is the problem of gauge[1-3].
We address this issue by considering a consistent perturbative expansion
of
the Einstein equations in first- and second-order perturbations of the
black hole metric. There are gauge transformations related to the
parametrization of the dynamical degrees of freedom in each order in
perturbation theory, and we use these symmetries judiciously to cast the
problem in as simple a manner as possible. The task of identifying
physical quantities in the effective energy-momentum tensor (henceforth
EEMT) of gravitational perturbations becomes much simpler that way.

The main point of this paper is the possibility of identifying physical
quantities in Regge-Wheeler gauge. This issue has also been raised
by Gleiser \cite{gleiser} for the case of the $\ell=2$, even-parity
gravitational perturbation. We explore the connections between the flux of
energy out of the black hole, contained in the $G^t_{\,r}$ equations, and
the inner product necessary both for the proper normalization of the
gravitational perturbations and for any attempts at quantizing the
gravitational fluctuations.

The outline of the paper is as follow: in Section 2 we review the
formalism connected with the perturbative expansion of the Einstein
equations to second order and define the EEMT of gravitational
perturbations, $T_{\mu \nu}^{GW}$. In Section 3 we write the linearized
perturbations and consider a scalar product that can fix their
normalizations. In Section 4 we give details about our first order gauge
choice: the Regge-Wheeler (RW) one. In Section 5 we show how the flux of
gravitational waves computed in the RW gauge satisfies all the physical
criteria of an energy flux both at infinity and on the horizon. In Section
6 we present a perturbative expansion of the physical degrees of freedom
(the Regge-Wheeler and Zerilli functions) that can be used in the
expressions for the energy, then we show that it is possible to normalize
these gravitational waves in the Schwarzschild background. Finally, we 
compute what that energy is in the case of the odd perturbations. We
conclude in Section 7.

\section{The General Plan}

We study back reaction by expanding the Einstein field equations to second
order in the initial values of the metric perturbations. The metric of
the perturbed black hole is given by the series

\be
g_{\mu \nu} = g_{\mu \nu}^{(0)} + \epsilon \g_{\mu \nu} 
+ \epsilon^2 g_{\mu \nu}^{(2)} + {\cal O} (\epsilon^3) \; , 
\label{secexp}
\ee
where $\epsilon$ is the perturbative parameter, which can be thought of as
the amplitude of the metric perturbation on the initial value surface. It is
useful for the moment to regard the first order metric perturbations as
gravitational waves propagating in the black hole background, and second
order metric perturbations as the response (back reaction) effected by the
gravity waves.

Both the gravity waves and their back reaction on the metric 
field can be
parameterized in an infinity of ways, reflecting the symmetry of the exact
theory under generic gauge transformations. Consistent with the expansion
above, we write gauge trasformations at first and second order in the form
\cite{AMB,bruni}

\begin{eqnarray}
\label{gt_1}
\widetilde{\g_{\mu \nu}} &=& 
\g_{\mu \nu} - {\cal L}_{\xi^{(1)}} g_{\mu\nu}^{(0)} \; , \\
\label{gt_2}
\widetilde{g_{\mu \nu}^{(2)}} &=& 
g_{\mu \nu}^{(2)} - {\cal L}_{\xi^{(1)}} \g_{\mu \nu} 
+ \frac{1}{2} {\cal L}_{\xi^{(1)}}^2 g_{\mu \nu}^{(0)} 
- \frac{1}{2} {\cal L}_{\xi^{(2)}} g_{\mu \nu}^{(0)} \; ,
\end{eqnarray}
where ${\cal L}$ is the Lie derivative and 
$\xi^{(1)\, \mu}$, $\xi^{(2)\, \mu}$ are two {\it independent} vectors. 
Their significance becomes clearer if we regard this gauge transformation
as being generated by a second order coordinate transformation:

\be
\label{gt_x2}
\tilde{x}^{\mu} = x^{\mu} + \epsilon \, \xi^{(1)\, \mu} + 
\frac{\epsilon^2}{2} \left( \xi^{(1)\, \mu}_{, \nu} \xi^{(1)\, \nu} 
+ \xi^{(2)\, \mu} \right) + {\cal O} (\epsilon^3) \; .
\ee
In practical terms, the statement is that the parametrization of the
first-order quantities (the gravitational waves) can be carried
independently of
the parametrization of the second-order quantities 
(the back reaction).\footnote{Note, however, that the parametrization 
at the first order does affect the second-order quantities through
nonlinear terms - e.g., the middle term on the right hand side of
(\ref{gt_2}).}

Back reaction is the feedback effect driven by the nonlinearities of the
Einstein field equations. At second order in the perturbations, we
expand Einstein's equations as
\be
\quad \quad  G_{\mu \nu}^{(1)} [g^{(2)}] + G_{\mu \nu}^{(2)} [\g] + \ldots 
=0 \; ,
\label{eps2}
\ee 
where the gravitational waves $\g$ obey the Einstein field equations in
vacuum,
\be
\quad \quad  G_{\mu \nu}^{(1)} [\g] = 0 \; .
\label{eps1}
\ee 
The notation should be clear: $G_{\mu \nu}^{(1)}$ is a differential 
operator which is linear in the arguments, and for this reason appears 
in Eq. (\ref{eps1}) acting on the first order metric perturbations, and 
in Eq. (\ref{eps2}) acting on the second order metric perturbations. 
$G_{\mu \nu}^{(2)}$ is a differential operator which is quadratic in its 
arguments.

We will consider Eq. (\ref{eps2}) for the spherically symmetric
(isotropic) second order perturbations. The Schwarzschild background
metric in $x = (t, r, \theta, \phi)$ coordinates is

\be 
\label{g_0}
g_{\mu \nu}^{(0)} = \mbox{diag} (-\G,\frac{1}{\G},r^2,r^2\sin^2\theta) \; , 
\ee
with $\G \equiv 1 - R/r$ and $R = 2 G M$. The metric to second order
is given by

\be 
g_{\mu \nu} = g_{\mu \nu}^{(0)}(r) + \epsilon \g_{\mu \nu} (x) 
+ \epsilon^2 g_{\mu \nu}^{(2)} (t, r) + {\cal O} (\epsilon^3) \; .
\label{secexp2} 
\ee 
We have omitted all second-order anisotropic terms from this
expression because at this order in perturbation theory they decouple
from the isotropic terms.

Note that by keeping the spherical symmetry in this problem does not imply
the time independence of the second order metric, which includes effects
from back-reaction: the time independence of the Schwarzschild metric is
the consequence of Birkhoff's theorem, which assumes no source terms for
the Einstein equations. Therefore, the metric of a black hole that
radiates can no longer be static.

The back-reaction equation we are interested in is the isotropic
projection of Eq. (\ref{eps2}), which we recast as:
\be
G_{\mu \nu}^{(1)} [g^{(2)}(t,r)] = 
- \langle G_{\mu \nu}^{(2)} [\g] \rangle_{\Omega} =
8 \pi G T_{\mu \nu}^{GW} \; ,
\label{eqav}
\ee
where the angle average is given by

\be
\langle F \rangle_{\Omega} = \frac{1}{4 \pi} 
\int d\theta \, d\phi \, F \; .
\ee
We call $T_{\mu \nu}^{GW}$ the effective energy-momentum tensor (EEMT) of
gravitational waves in the Schwarzschild background.

The gauge transformation at the second order (\ref{gt_2})
is rewritten as:
\be
\widetilde{g_{\mu \nu}^{(2)}}(t, r) =
g_{\mu \nu}^{(2)}(t, r) - \langle {\cal L}_{\xi^{(1)}} \g_{\mu \nu}
\rangle_{\Omega} -
\frac{1}{2} {\cal L}_{\tilde{\xi}^{(2)}} g_{\mu \nu}^{(0)} +
\frac{1}{2} \langle {\cal L}_{\xi^{(1)}}^2 g_{\mu \nu}^{(0)}
\rangle_{\Omega} \,,
\label{gaugeav}
\ee
where now $\tilde{\xi}^{(2)}_\mu = [ \tilde{\xi}^{(2)}_t (t, r) ,
\tilde{\xi}^{(2)}_r (t,r) , 0 , 0 ] $, i.e.
only the gauge freedom according to the spherical symmetry
remains. It will be clear in the next sections, when
we explicitly consider the expansion of the perturbations
in spherical harmonics, that Eqs. (\ref{eqav}) and (\ref{gaugeav})
follow from the averaging of Eqs. (\ref{eps2}) and (\ref{gt_2}) respectively
-- since the second order quantities appear
linearly in the second order relations, only the monopole part
of $g_{\mu \nu}^{(2)}$ and
$\xi^{(2)}_\mu$ survive to the averaging, and these are precisely
$g_{\mu \nu}^{(2)}(t, r)$ and $\tilde{\xi}^{(2)}_\mu$.
The monopole part of the metric perturbation contains four degrees of
freedom, but we can use the two components of $\tilde{\xi}^{(2)}_\mu$ to
cancel two of the degrees of freedom, so we are left with two independent
degrees of freedom. This is the right number of functions needed to
describe a time-dependent isotropic 
metric, as for the case of an evaporating black hole \cite{bardeen}.

Finally, let us comment on three basic points. 

First, it is easy to verify that the EEMT obeys conservation equations
which are just the Bianchi identities expanded to second order in
perturbation
theory. These conservation equations are completely independent of the
form of the second-order degrees of freedom, since the part of the
Bianchi identities which is linear in $g_{\mu\nu}^{(2)}$ and its
derivatives is zero identically (see also \cite{PRD1999}).

Second, we do not address the problem of the
gauge dependence (or independence) of the EEMT of gravitational waves.
Once one has fixed his gauge choices at the first and second
order, one can then try to extract physical information from the analysis 
of the second order Einstein equations or from considering 
gauge-invariant observables.

Third, in what follows we consider the problem of linearized perturbations
in vacuum. Classically, we would not have any physical solutions
without source terms. Here we could think of the gravitational
waves as having been generated by quantum effects close to the horizon
(Hawking evaporation \cite{hawking}).

\section{Linearized perturbations in vacuum}

Consider the Einstein field equations in vacuum, linearized  
around a background geometry $g_{\mu \nu}^{(0)}$:

\begin{eqnarray}
G_{\mu \nu}^{(1)} &=&
\g_{\mu \nu \, |\alpha}^{\,\,\,\,\quad|\alpha} 
- \g_{\mu \alpha | \nu}^{ \,\,\,\,\quad|\alpha}
- \g_{\nu \alpha |\mu}^{ \,\,\,\,\quad|\alpha}
+ \g_{|\nu \, |\mu} 
+ g_{\mu \nu}^{(0)} \left( \g_{\alpha \beta}^{|\alpha \, |\beta} 
- \g_{|\alpha}^{\quad |\alpha} \right) \nonumber \\
&&
+ \g_{\mu \nu} R^{(0)} - g_{\mu \nu}^{(0)} \g_{\alpha \beta} R^{(0)\, \alpha 
\beta} = 0
\end{eqnarray}
where $\g$ is the trace of $\g_{\mu \nu}$ and $|$ denotes the covariant
derivative with respect to the background metric. By subtracting
from the above equation its trace and considering Ricci flat background
spacetimes (such as Schwarzschild) we have

\be 
\g_{\mu \nu |\alpha}^{\,\,\,\,\quad|\alpha} - \g_{\mu \alpha |\nu}^{
\,\,\,\,\quad|\alpha} - \g_{\nu \alpha |\mu}^{ \,\,\,\,\quad|\alpha} +
\g_{|\nu\mu} = 0 \; , 
\label{lin1} 
\ee 
and 

\be 
\g_{\alpha \beta}^{\quad |\alpha\beta} - \g_{|\alpha}^{\, \,
\,|\alpha} = 0 \; .
\ee
It is easy to see that the last equation is trivial if one uses
the transverse ($\g_{\alpha \beta}^{\quad |\alpha} = 0$) and traceless
($\g = 0$) gauge. By using the relation 

\be \g_{\mu \alpha | \nu}^{
\,\,\,\,\quad|\alpha}=\g_{\mu\alpha \,\,\,\, |\nu}^{ \quad |\alpha} 
+ R^{\alpha \quad \beta (0)}_{\,\, \mu \nu} \g_{\alpha
\beta} + R^{\alpha (0)}_{\,\, \nu} \g_{\mu \alpha} \, 
\ee 
we can switch the order of derivatives in Eq. (\ref{lin1}):

\be 
\label{EOM_2}
\g_{\mu \nu \,|\alpha}^{\,\,\,\,\quad| \alpha} - 
\g_{\mu \alpha \,\,\,\, | \nu}^{\quad |\alpha} - 
\g_{\nu \alpha \,\, |\mu}^{ \quad |\alpha} +
2 R^{\alpha \,\,\beta \, (0) }_{\,\,\, \mu \,\,\, \nu}
\g_{\alpha \beta} + \g_{|\nu\mu} = 0 \; .  
\ee

We can associate a scalar product with these equations of motion
\cite{candelas}:

\begin{eqnarray}
\label{norm}
\langle \psi^{\alpha \beta}, \phi_{\alpha \beta} \rangle &=& 
-i \int_{\Sigma} d^3 x \sqrt{g_{{}_\Sigma}} n^{\mu} 
\\ \nonumber
&& \times \left[ 
\psi^{\alpha \beta \, *} \bar{\phi}_{\alpha \beta |\mu} - 
\phi^{\alpha \beta} \bar{\psi}_{\alpha \beta | \mu}^{*} - 
2 \left( 
\bar{\psi}_{\mu \nu}^* \bar{\phi}^{\alpha \nu}_{\quad |\alpha} - 
\bar{\phi}_{\mu \nu} \bar{\psi}^{\alpha \nu \, *}_{\quad |\alpha} 
\right) 
\right] \; , 
\end{eqnarray}
where $\psi$ and $\phi$ are two generic (complex) solutions to the
equations of motion (\ref{EOM_2}), $n^{\mu}$ is the unit vector normal to
the
spacelike hypersurface $\Sigma$, $g_{{}_\Sigma}$ is the determinant 
of the induced metric on $\Sigma$ and $\bar{\psi}_{\mu \nu}$ is defined by

\be 
\bar{\psi}_{\mu \nu} = \psi_{\mu \nu} - \frac{1}{2} g_{\mu \nu}^{(0)}
\psi^\alpha_\alpha \; .
\ee 

The normalization of a gravity wave mode should be given by the inner
product
(\ref{norm}). However, because metric perturbations diverge at 
infinity in the RW gauge, it
is common practice to neglect the Klein-Gordon inner product (\ref{norm})
and instead to normalize the gravity waves through their mass-energy in an
asymptotically flat region of space. Since we would like to compute the
mass-energy of the gravitational waves from first principles, we will
avoid this procedure. Instead, we show how to obtain asymptotic
expressions for the gravitational waves in Regge-Wheeler gauge, and use
them in the inner product (\ref{norm}), which then becomes well-defined.

\section{The Regge-Wheeler gauge}

Under a gauge transformation (\ref{gt_x2}), the metric perturbations are
transformed according to the general laws (\ref{gt_1}) and (\ref{gt_2}). 
In order to choose a gauge we must specify constraints for the metric
components that uniquely fix the eight free functions $\xi^{(1)\mu}$ and
$\xi^{(2)\mu}$ with respect to an arbitrary gauge. The penalty for not
fixing the gauge completely is having to deal with unphysical degrees of
freedom (which become ghosts after quantization),
corresponding to the free functions that were left unconstrained.

To first order in perturbation theory, one of the (infinitely many) 
choices of coordinates which fixes the gauge completely in the
Schwarzschild background is the Regge-Wheeler 
gauge\cite{regge,zerilli,pullin}. 
This choice of gauge is exceptionally convenient, since the tensor
structure of the Einstein field equations decouple from their angular
dependence. The two degrees of freedom of gravitons in vacuum correspond
to two (orthogonal) sets of perturbations, odd and even (or electric and
magnetic), depending on how they transform under parity ($\vec{x}
\rightarrow -\vec{x}$).

In the spherically symmetric background (\ref{g_0}), the perturbations are 
expanded in spherical harmonics $Y_{\ell}^{m}(\theta,\phi)$ and therefore 
the modes carry multipole numbers $(\ell,m)$. In Regge-Wheeler gauge we
have, for the odd metric perturbations:

\be
\g_{\mu \nu}^{o} = 
  \left(\begin{array}{cccc} 
	0 	& 	
	0 	&  
	- \frac{{h}_{0}^{\ell m}}{\sin \theta}  
	{\partial Y_\ell^{m}\over \partial \phi} 	& 
	{h}_{0}^{\ell m}{\sin \theta}
	{\partial Y_\ell^{m}\over \partial \theta} 
  \\ 
 	0 	& 
	0 	&
	- \frac{{h}_{1}^{\ell m}}{\sin \theta}   
	{\partial Y_\ell^{m}\over \partial \phi} 	&
	{h}_{1}^{\ell m}{\sin \theta}
	{\partial Y_\ell^{m}\over \partial \theta}   
  \\ 
	\mbox{sym} 	&  
	\mbox{sym} 	&
	\mbox{0}	&
	\mbox{0}
  \\ 
	\mbox{sym} 	&  
	\mbox{sym} 	& 
	\mbox{0}	&
	\mbox{0}	
\end{array}\right)\, 
\label{odd}
\ee
where ${h}_{0}^{\ell m}$ and ${h}_{1}^{\ell m}$ are functions of
$t$ and $r$.

The even parity perturbations are given by 
\be
\g_{\mu \nu}^{e} =  
  \left(
	\begin{array}{cccc}  
	\G H_{0}^{\ell m} Y_\ell^{m} 	& 
	L^{\ell m}Y_\ell^{m} 	& 
	\mbox{0}	&
	\mbox{0}
  \\ 
	\mbox{sym} 	& 
	\frac{H_{2}^{\ell m}}{\G}  Y_\ell^{m}	&
	\mbox{0}	&
	\mbox{0}
  \\
	\mbox{0} 	& 
	\mbox{0} 	& 
	r^2 K^{\ell m} Y_{\ell}^m	&
	\mbox{0} 
  \\
	\mbox{0} 	& 
	\mbox{0} 	& 
	\mbox{0} 	&
	r^2 K^{\ell m} Y_{\ell}^m \sin^2 \theta	
  \end{array} 
  \right) \,  
\label{even}
\ee 
where $H_{0}^{\ell m}$, $L^{\ell m}$, $H_{2}^{\ell m}$ and $K^{\ell m}$
are also functions of $t$ and $r$.

With the metric perturbations written in this form, the equations of
motion can be separated and cast in the simple form (see for example
\cite{regge,zerilli,pullin})

\be
\label{chieq}
\frac{\partial^2 \chi^{o,e}} {\partial t^2}  
- \frac{\partial^2 \chi^{o,e}}{\partial r^{*\, 2}} 
+ V^{o,e} (r) \chi^{o,e} = 0 \; ,
\ee
where the superscripts $o,e$ correspond to the odd or even perturbations
and $r^*$ is the so-called tortoise coordinate,

\be
\label{tortoise_r}
r^* = r + R \ln \left ( \frac{r}{R} - 1 \right) \; .
\ee
The odd and even potentials $V^{o,e}$ are given respectively by:

\be
\label{v_odd}
V^{o} (r) = \G \left[ \frac{\ell(\ell+1)}{r^2} 
- \frac{3 R}{r^3} \right]
\ee
and

\be
\label{v_even}
V^{e}(r) = \G \frac{ 2 \lam^2 ( \lam +1) r^3 + 3 \lam^2 R r^2 
+ \frac{9}{2} \lam R^2 r + \frac{9}{4} R^3}
{r^3 ( \lam r + \frac{3}{2} R)^2} \; , 
\ee
where $\lambda \equiv \ell (\ell+1)/2 -1$.

Both the even and odd potentials have the same asymptotic form in the
limit $r \rightarrow + \infty$:

\be
V^{o,e}(r) \simeq 2 \frac{\lam + 1}{r^2} \,.
\ee
Both also go to zero on the horizon, but with a different slope:

\be
V^{o}(r) \simeq \G \frac{2 \lam - 1}{R^2} \; , \quad 
V^{e}(r) \simeq \frac{2\G}{R^2} 
\frac{\lam^2 + \lam + 3/4}{\lam + 3/2} \; .
\ee

The two odd functions ${h}_{0}^{\ell m}$ and $h_{1}^{\ell m}$ are easily
expressed in terms of the Regge-Wheeler function $\chi^o$:

\begin{eqnarray}
\frac{\partial {h}_{0}^{\ell m}}{\partial t} &=& 
\frac{\partial}{\partial r^*} ( r \chi^o )  
\nonumber \\
h_{1}^{\ell m} &=& \frac{r}{\G} \chi^o \; .
\label{oddrel}
\end{eqnarray}

The even metric perturbations can be written in terms of the Zerilli
function $\chi^e$ as well. Since one of the linear equations of motion
states that $H_2^{\ell m} = H_0^{\ell m} \equiv H^{\ell m}$, there
are only three different nonzero metric components:

\begin{eqnarray}
K_{\ell m} &=& f_\ell (r) \chi^e + \frac{\partial \chi^e}{\partial r^*}  
\nonumber \\ 
\label{evenrel}
L_{\ell m} &=&  \frac{\partial}{\partial t}   
\left[  g_\ell (r) \chi^e + \frac{r}{\Gamma} \frac{\partial
\chi^e}{\partial r^*} \right]
\\
H_{\ell m} &=& {\partial {\; } \over \partial r} 
\left[ \G g_\ell (r)  \chi^e   + r \G {\partial \chi^e \over \partial r} 
\right]  -  K \nonumber \,,
\end{eqnarray}
where

\begin{eqnarray}
\label{f}
f_\ell (r) &\equiv& 
\frac{ \lam (\lam + 1) r^2 + 3/2 Rr \lam + 3/2 R^2} {r^2 (\lam r + 3/2 R)}
\; ,
\\ 
\label{h}
g_\ell (r) &\equiv& \frac{1}{r \G} 
\frac{\lam r^2 - 3/2 \lam R r -3/4 R^2}{\lam r + 3/2 R} \; .
\end{eqnarray}

Some remarks are in order. Note that the potentials $V^{o,e}(r)$
vanish asymptotically both on the horizon and at infinity ($r^*
\rightarrow \mp \infty$). On these asymptotic regions the solutions to the
wave equation (\ref{chieq}) can be expanded in terms of plane waves in
retarded or advanced time,

\be
\chi^{o,e} \sim e^{- i \omega ( t \pm r^*)} \; ,
\label{oddplane}
\ee
where the plus holds for ingoing (towards the horizon) waves and the
minus holds for outgoing waves.

This implies that in Regge-Wheeler gauge the odd and even metric
perturbations diverge both on the horizon and at infinity: for example,
$h_1 \simeq r^* \chi^o$ when $r^* \rightarrow \infty$ and $h_1 \simeq R
e^{-r^*/R} \chi^o$ when $r^* \rightarrow -\infty$. Of course, this is a
coordinate artifact of the Regge-Wheeler gauge, and physical quantities
should remain finite in that (or any other) gauge, as long as we keep away
from the singularity at the center of the black hole. There are gauges in
which physical quantities are manifestly regular at infinity, such as
asymptotically flat \cite{pullin} or radiation gauge \cite{zerilli}. 
Our purpose is to show (see Section 5) that the
Regge-Wheeler gauge can also be used to compute physical quantities, even
if the metric components diverge.

We can simplify greatly the ansatz for the metric if we adopt the standard
procedure of rotating the $z$-axis to put each mode $(\ell,m)$ in the
state $(\ell,0)$. The $\phi$ dependence thus drops out of the metric
perturbations (\ref{odd})-(\ref{even}), and all subsequent
equations involve only Legendre polynomials $P_{\ell} (\cos\theta)$ and
their derivatives. Of course this simplification does not change 
our results for the back reaction on the isotropic mode.

If the metric ansatz (\ref{odd})-(\ref{even}) is appropriate to describe
the anisotropic metric perturbations around the black hole, what is
appropriate to describe the back reaction of these perturbations on the
isotropic background? Physics dictates that we should expect some mass to
be lost by the black hole. However, the mass appears both in $g_{00}=
-(1-2GM/r)$ and in $g_{11} = (1-2GM/r)^{-1}$. Generically there should be
two degrees of freedom describing the (no longer static)
isotropic metric\cite{bardeen}, which we define adopting the following
gauge choice at the second order:

\begin{eqnarray}
\label{br00}
g_{00} &\rightarrow& - 1 + \frac{2GM}{r} - \epsilon^2 
\frac{2G\Delta_0 M(r,t)}{r} + {\cal O} (\epsilon^3) \; ,\\
\label{br11}
g_{11} &\rightarrow& \left[ 1 - \frac{2GM}{r} + \epsilon^2
\frac{2G\Delta_1 M(r,t)}{r} \right]^{-1} + {\cal O} (\epsilon^3) \; .
\end{eqnarray}

The interpretation of $\Delta_0 M$ is related to the ADM mass at infinity,
while $\Delta_1 M$ is related to the location of the horizon\footnote{We
thank Roberto Casadio for having pointed out this feature to us.}.
However, sometimes it is useful to make the following approximation:
assume that
the black hole emits radiation in packets, such that for a spherical shell
of radius $r=\bar{r}$ from the black hole, any given packet is either
completely
inside or outside the shell. If the gravity waves packet is inside the
shell, the metric describing the isotropic background is just the exterior
Schwarzschild metric without corrections. If the packet is outside the
shell, the mass of the black hole measured by an observer located at 
$r=\bar{r}$ has changed by an amount $\Delta M$ that
is now independent of time and radius. In the limit of this
approximation we can ignore subtleties with the dynamics that may
distinguish $\Delta_0 M(r,t)$ from
$\Delta_1 M(r,t)$ in a more complex physical situation.

Let us comment on a peculiarity of the gauge 
choice at second order. There is no way of fixing completely the 
gauge for the isotropic perturbation, since the remaining freedom is related 
to a redefinition of time:
\be
t \rightarrow t + \frac{\epsilon^2}{2} f(t) + {\cal O} (\epsilon^3)
\ee
This is also related to the fact that time is defined {\em ad hoc} 
for the Schwarzschild solution \cite{mtw} (which for us is the 
zero order metric of our perturbative analysis).

\section{The flux of gravitational waves}

In this section we will focus on the mixed $t-r$ component of the EEMT of
gravitational waves for two main reasons. The first is simplicity:
$G^t_{\; r}=R^t_{\; r}$ to second order has a short expression in terms of
the perturbations. The second is that this term represents an energy flux.
Consider for example a scalar field in the Schwarzschild background. The
mixed $t-r$ component of the scalar's energy-momentum tensor is given by: 

\begin{eqnarray}
T^t_{\; r}  = \frac{1}{\G^2} \varphi,_{t} \varphi,_{r^*} ( 2 \xi - 1) 
+ 2 \frac{\xi}{\G^2} \varphi \left( \varphi,_{tr^*} - \frac{R}{2r^2}  
\varphi,_{t} \right)
\; ,
\label{scalar}
\end{eqnarray}
where $\xi$ is the coupling of the scalar field to the curvature.
By considering the proper normalization $\sim 1/r$ in front of the
asymptotic spherical plane waves solution for a scalar field
one has
an asymptotic behaviour ${\cal O}(1/\G^2)$ on the
horizon and ${\cal O}(1/r^2)$ at infinity. We observe that the behaviour
of $T^t_{\; r}$ on the horizon is not really pathological, indeed it is
regular in a freely falling frame.

We now proceed to calculating the mixed $t-r$ component of the effective
energy-momentum tensor (EEMT) of gravitational perturbations around a black
hole. Since the even and the odd degrees of freedom are orthogonal to each
other at this order in perturbation theory, we can consider their contributions
to the EEMT separately. For the choice of second order metric 
coefficient in Eq. (\ref{br11}) one has: 
\be
R^{t \, (1)}_{\; r} [g^{(2)}(t,r)] = 
- \frac{2G\Delta_1 \dot{M}}{r^2\G}
\ee

The odd contribution to the $t-r$ component in Regge-Wheeler gauge is,
after averaging over angles:

\begin{eqnarray}
\label{T_01_odd}
\langle R^t_{\; r \, odd} \rangle_\Omega = 
	\sum_\ell \frac{\ell(\ell+1)}{2\ell+1} 
	\frac{h_1}{2 r^4} 
	\left[ 
	r^2 (\dot{h}_{1}' - h_{0}'' ) 
	- 2 r \dot{h}_{1} + 2 h_0 \right] \; ,
\end{eqnarray}
where a dot and a prime denote derivatives with respect to $t$ and $r$ 
respectively.

The first term inside square brackets in (\ref{T_01_odd})  appears to go
as $r^3$ at $r \rightarrow \infty$ (since $h_0 , h_1 \propto r$ in this
limit). If that was so, the radiated mass $\Delta M$ would be divergent. 
In reality, that term is at most $\propto r^2$ because the leading term
cancels upon use of relations (\ref{odd}) and the equations of motion
(\ref{chieq}). Nevertheless, the ${\cal O}(1/r)$ term 
in $\langle R^t_{\; r \, odd} \rangle_\Omega$ still gives a
contribution that could make $\Delta M$ divergent. We show in the next
section that these apparently divergent terms in Regge-Wheeler gauge are
similar to the terms in the normalization of the gravity waves that are
also apparently divergent. However, in both cases the divergences are
revealed to be fictious, and the physical information can be retrieved. 

On the horizon the leading term in the right hand side in (\ref{T_01_odd})
appears to go as ${\cal O} (1/\Gamma^3)$. Also in this case by using the 
equation of motion (\ref{chieq}) this term vanishes, and 
so $T^{t\, GW}_{\; r} \sim {\cal O} (1/\Gamma^2)$, as in the scalar field 
case (\ref{scalar}).

The even contribution to the $t-r$ mixed component has a similar structure:

\begin{eqnarray}
\label{T_01_even}
\langle R^t_{r\, even} \rangle _\Omega &=& 
	\sum_{\ell} \frac{1}{2\ell+1} 
	\frac{1}{r^2 \G} 
	\left[ 
	2 r^2 \dot{K}' (K-H) + 2 r^2 \G K''L 
	\right. 
	\\ \nonumber 
	&& 
	+ r^2 K' (\dot{K}+\dot{H})
	- r^2 \dot{K} H' 
	+ 4 r \G K' L 
	\\ \nonumber
	&&
	+ \frac{2 r \dot{K}}{\G} (1-\frac{3GM}{2r})(K-H)
	- \left. \ell(\ell+1) KL \right] \; .
\end{eqnarray}
Again, from (\ref{evenrel}) there is an apparent divergence in 
$T^{t\, GW}_{\; r}$ at infinity which is  given by terms ${\cal O} (r^3)$ 
inside the square brackets in (\ref{T_01_even}). 
However, the leading term vanishes by using (\ref{chieq}), and 
the physical information can be obtained among the 
subleading terms.  Also in this case the apparent divergence on the
horizon ${\cal O} (1/\Gamma^3)$ vanishes by using (\ref{chieq}). 
\footnote{ Note that all these cancellations of the most divergent terms 
by using the equations of motions are also a property of the energy 
momentum tensor of gravitational waves, even before the average over angles.}

\section{Asymptotics of the Regge-Wheeler and Zerilli functions}

Consider now the normalization of the metric perturbations in
Regge-Wheeler gauge. Since perturbations with different parity decouple
the odd contribution to the inner
product (\ref{norm}) can be written as:

\begin{eqnarray}
\label{norm_odd}
\langle \psi^{\alpha \beta}, \phi_{\alpha \beta} \rangle_{odd} &=& 
	-i \sum_\ell \frac{4 \ell(\ell+1)}{2\ell+1} \int_\Sigma 
	dr^* \G 
\\ \nonumber
&&	\times \left[ h_1 \dot{h}_1^* - h_1^* \dot{h}_1
	+ \frac{h_0^*}{r} (r h_1'+ 2 h_1) 
	- \frac{h_0}{r} (r {h_1^*}'+ 2 h_1^*)
	\right] \; ,
\end{eqnarray}
where the angle integrations have already been performed.
When $r \rightarrow \infty$ the integrand appears to be $\propto r^2$.
This structure appears in the the even contribution to the inner product 
as well.

We show next that in fact no divergences survive in the normalization
conditions once the perturbative solutions to the wave equation
(\ref{chieq}) are used in the expression above.

From Eq. (\ref{chieq}) for the Zerilli ($\chi^e$) and Regge-Wheeler
($\chi^o$) functions it is clear that at infinity they assume the form of
a plane wave in retarded time $u=t-r^*$ (or advanced time $v$, in the case
of
an incoming mode). This suggest an expansion of $\chi$ in terms of
functions of both retarded time and radius of the form\cite{pullin}

\be
\label{exp_chi}
\chi(r,t) = X_0(u) + r^{-1} X_1 (u) + r^{-2} X_2 (u) 
	+ {\cal O}(r^{-3})  \; .
\ee
This expansion should be consistent as long as the gravity wave packet
does not probe the region where the potential is large, $r \approx 3GM$. 
In other words, the approximation breaks down as soon as the
reflection and transmission of the waves become important and the very 
description of a wave packet in terms of purely ingoing (or
outgoing) modes breaks down.

We can substitute the ansatz above into the wave equation (\ref{chieq}) and
solve the hierarchy of equations that ensue. Since the leading order terms
in the potentials $V^o$ and $V^e$ are the same to order $r^{-2}$, 
the solutions for $\chi^e$ and $\chi^o$ are identical to each other at 
that order. The result is, 
for an ingoing gravity wave of even or odd parity,

\begin{eqnarray}
\label{sol_chi}
\chi^{o,e}(r,t) &=& \frac{1}{\lambda + 1} \ddot{X}^{e,o}(u)
	+ r^{-1} \dot{X}^{e,o} (u) 
\\ \nonumber
&&	+ r^{-2} \left[ 
		\frac{\lambda}{2} X^{e,o}(u)
		- \frac{3GM}{2(\lambda + 1)} \dot{X}^{e,o}(u)
		\right] + {\cal O}(r^{-3}) \; .
\end{eqnarray}
Notice that $X$ is a completely
{\it generic} function of retarded time, not 
necessarily a plane wave, reflecting the fact that we have not yet imposed
any boundary conditions on the problem. We also stress that $X$ carries
an index $\ell$, like the mode of metric perturbations.

From now on we focus on the odd perturbations for the sake of simplicity,
since the expressions for the even perturbations are too cumbersome, and
the physics and the lessons we draw are precisely the same as in the odd
case. With the help of relations (\ref{oddrel}) we can express the metric
perturbations in terms of the odd function $X(u)$: 

\begin{eqnarray}
\label{h0_X}
h_0 &=& - \frac{r}{\lambda + 1} \ddot{X} - \frac{\lambda}{\lambda + 1}
\dot{X} 
	+ r^{-1} \left( -\frac{GM}{2(\lambda + 1)} \dot{X} -
\frac{\lambda}{2} X \right) 
	+ {\cal O} (r^{-2}) \; , \\
\label{h1_X}
h_1 &=&  \frac{r}{\lambda + 1} \ddot{X} 
	+ \left( \frac{2GM}{\lambda + 1} \ddot{X} + \dot{X} \right) 
\\ \nonumber
&&	+ r^{-1} \left( -\frac{4 G^2M^2}{\lambda + 1} \ddot{X} 
	+ GM \frac{4 \lambda + 1}{2(\lambda + 1)} \dot{X} +
\frac{\lambda}{2} X \right) 
	+ {\cal O} (r^{-2}) 
\; .
\end{eqnarray}

It is now a matter of algebra to substitute these expressions into
Eqs. (\ref{T_01_odd}) and (\ref{norm_odd}). The result for the
normalization is

\be
\label{norm_X}
\langle \psi^{\alpha \beta}, \phi_{\alpha \beta} \rangle_{odd}
= i \frac{8(\lambda + 1)}{2\ell+1} 
	\int_\Sigma \frac{dr^*}{\G} 
	\left[	r \frac{\lambda}{(\lambda + 1)^2} \frac{\partial}{\partial
t} 
	(\dot{X_\psi} \ddot{X}_\phi^*) + {\cal O} (r^0) \right] 
	\; .
\ee
Therefore, the ${\cal O} (r^3)$ divergence in the integral
(\ref{norm_odd}) has simply
disappeared after we used the asymptotic formula (\ref{sol_chi}).

The first term in the expression above would make the integral diverge as 
$r^2$. However, that term vanishes because it is the time derivative of 
an integral over the spacelike hypersurface $\Sigma$ (our assumptions are 
that the gravitational wave packet is either entirely inside or
entirely outside $\Sigma$.)
There is a physically intuitive way to argue that this 
divergent term should vanish. In the WKB approximation the $X_a$'s are 
plane waves, $X_a \propto \exp (-i w_a u)$. 
In this framework, the phases in the integrand of (\ref{norm_X}) would
interfere destructively.

In either case, the normalization of the odd metric perturbations is given
by the next term in perturbation theory:
\be
\label{norm_X_rad}
\langle \psi^{\alpha \beta}, \phi_{\alpha \beta} \rangle
_{\rm odd, WKB} 
	= i \frac{24 \lambda^2}{(\lambda+1)(2\ell+1)} 
	\int_\Sigma dr^* 
	( \dot{X}_\psi \ddot{X}_\phi^* - 
	  \dot{X}_\phi \ddot{X}_\psi^* ) 
\; .
\ee
If the mode solutions $X_\psi$ and $X_\phi$ are just WKB modes,
the integral above gives the expected normalization \cite{candelas}
with a delta function over the frequencies of the modes, $\delta(w_\psi -
w_\phi)$.

Therefore, it is possible to normalize gravitational
waves in Schwarzschild background using the inner product 
in Regge-Wheeler gauge. As we show below, a similarly 
divergent time derivative term appears in the equation for
the time variation of the mass of the black hole.
Again, the next to leading term is the finite one, in agreement with a
finite gravitational radiation flux from an object with finite mass. 
We note that the term in (\ref{norm_X}) which is proportional to $r$
at infinity contains also a divergence on the horizon: we have just 
shown how deal with it. The next term in (\ref{norm_X_rad}) is regular 
also on the horizon, since an expansion in powers of $\Gamma$ 
similar to (\ref{exp_chi}) holds close to the horizon.

Let us consider now the flux of energy given by Eq. 
(\ref{T_01_odd}). Substituting the expressions (\ref{h0_X}) and (\ref{h1_X})
we obtain the following (apparently) divergent term:

\be
\label{flux_div}
\Delta_1 \dot{M} = \sum_\ell \frac{2}{(\lambda+1)(2\ell+1)} r 
	\frac{\partial}{\partial t} \ddot{X} \ddot{X}^* 
	+ {\cal O} (r^0)  \; .
\ee

Some authors have assumed\cite{pullin,gleiser} that
$X$ obeys some asymptotic conditions such that the ${\cal O}(r)$ term
vanishes. It appears to us that the WKB approximation could already be
sufficient to ensure the finiteness of the radiated mass.

The radiated mass from odd-parity gravitational waves comes
from the next order term, and using (\ref{h0_X})-(\ref{h1_X}) 
the final result is

\be
\label{rad_mass_odd}
\Delta_1 M = \sum_\ell \frac{2(\lambda+1)^2}{(2\ell+1)} 
	\int \, dt \, \left( \chi^o \right)^2 
	+ {\cal O} (r^{-1}) \; ,
\ee
which is the expression that has been known for a long time \cite{cpm} 
and was first found using the Landau-Lifshitz pseudo-tensor
in the asymptotically flat gauge. Other authors \cite{pullin,gleiser} 
have obtained similar results for the even perturbations
(in the $\ell=2$ multipole case), also using the Regge-Wheeler
gauge.

\section{Conclusions}

We have explored the possibility of performing back-reaction computations
for the Schwarzschild black-hole in the Regge-Wheeler gauge. 
The Regge-Wheeler choice has the advantage of fixing completely 
the gauge, but it has the disadvantage of divergencies in the 
metric coefficients both on the horizon and at spatial infinity.

We have shown that divergencies in the gravitational energy 
flux at infinity and on the horizon, which could
arise from choosing the Regge-Wheeler gauge, are in fact nonexistent. 
Furthermore, the same types of cancellations 
arise in the evaluation of the inner product, which is
related to the equations of motion of the linear perturbations. 
This could represent a viable check for the amplitude of 
gravitational waves, independent from the requirement that the energy flux 
of gravitational waves must be finite at infinity. The investigation of
the effect of the back-reaction of gravitational radiation on the
horizon is under current investigation.

\vskip 0.3cm

\noindent {\bf Acknowledgments}

\vskip 0.3cm

\noindent We would like to thank R. Brandenberger 
for having suggested the problem and for useful suggestions during all 
the stages of the work. We would like to thank R. Balbinot for
several useful conversations. We also thank J. Pullin for making his notes
available to us. F. F. thanks W. Unruh and R. Wald for 
discussions. R. A. would like to thank the Physics Department of the
University of Bologna, where part of this work was completed, for its
hospitality. This work has been partially supported by the U.S. DOE under
Contract DE-FG0297ER41029, Task A (R. A.)  and by the INFN (F. F.).

\end{document}